\documentclass{ieeeaccess}

\usepackage{graphicx}
\usepackage{amssymb}
\usepackage{amsmath}
\usepackage{gensymb}
\usepackage{cite}
\usepackage{amsmath,amssymb,amsfonts}
\usepackage[export]{adjustbox}
\usepackage{algorithmic}
\usepackage{graphicx}
\usepackage{textcomp}
\usepackage{verbatim}
\usepackage{array}
\IEEEoverridecommandlockouts
\usepackage{listings}
\usepackage{tabularx}
\newcolumntype{b}{X}
\newcolumntype{s}{>{\hsize=.5\hsize}X}

\def\BibTeX{{\rm B\kern-.05em{\sc i\kern-.025em b}\kern-.08em
		T\kern-.1667em\lower.7ex\hbox{E}\kern-.125emX}}
\begin{document}
	\history{Date of publication xxxx 00, 0000, date of current version xxxx 00, 0000.}
	\doi{10.1109/ACCESS.2022.DOI}
	
	\title{Adaptive Pulse Compression for Sidelobes Reduction in Stretch Processing based MIMO Radars}
	\author{\uppercase{Hamza Malik}\authorrefmark{1}, 
		\uppercase{Jehanzeb Burki\authorrefmark{2}, and Muhammad Zeeshan Mumtaz}\authorrefmark{3}}
	\address{College of Aeronautical Engineering, National University of Sciences and Technology, Pakistan}
	\address[1]{hmalik@cae.nust.edu.pk}
	\address[2]{jehanzeb.ahmad@gatech.edu,  jehanzeb.ahmad@cae.nust.edu.pk} \address[3]{zmumtaz@cae.nust.edu.pk}

	\markboth
	{Hamza \headeretal: Adaptive Pulse Compression for Sidelobes Reduction in Stretch Processing based MIMO Radars}
	{Hamza \headeretal: Adaptive Pulse Compression for Sidelobes Reduction in Stretch Processing based MIMO Radars}
	
	\corresp{Corresponding author: Hamza Malik (e-mail: hmalik@cae.nust.edu.pk).}
	
	\begin{abstract}
		Multiple-Input Multiple-Output (MIMO) radars provide various advantages as compared to conventional radars. Among these advantages, improved angular diversity feature is being explored for future fully autonomous vehicles. Improved angular diversity requires use of orthogonal waveforms at transmit as well as receive sides. This orthogonality between waveforms is critical as the cross-correlation between signals can inhibit the detection of weaker targets due to sidelobes of stronger targets. This paper investigates the Reiterative Minimum Mean Squared Error (RMMSE) mismatch filter design for range sidelobes reduction for a Slow-Time Phase-Coded (ST-PC) Frequency Modulated Continuous Wave (FMCW) MIMO radar. Initially, the performance degradation of RMMSE filter is analyzed for improperly decoded received pulses. It is then shown mathematically that proper decoding of received pulses requires phase compensation related to any phase distortions caused due to doppler and spatial locations of targets. To cater for these phase distortions, it is proposed to re-adjust the traditional order of operations in radar signal processing to doppler, angle and range. Additionally, it is also proposed to incorporate sidelobes decoherence for further suppression of sidelobes. This is achieved by modification of the structured covariance matrix of baseline single-input RMMSE mismatch filter. The modified structured covariance matrix is proposed to include the range estimates corresponding to each transmitter. These proposed modifications provide additional sidelobes suppression while it also provides additional fidelity for target peaks. The proposed approach is demonstrated through simulations as well as field experiments. Superior performance in terms of range sidelobes suppression is observed when compared with baseline RMMSE and traditional Hanning windowed range response.
	\end{abstract}
	
	
	\titlepgskip=-15pt
	
	\maketitle
	
	\section{Introduction}
	\label{sec:introduction}
	\PARstart{M}{IMO}  radar requires the use of waveforms which are orthogonal both at transmit and receive sides. As long as these orthogonal waveforms can be separated at the receive side, only then the angular diversity property related to MIMO radars can be implemented \cite{li2008mimo}.  The orthogonality requirement ensures minimum interference between transmitted waveforms, provided they can be separated in the receiver using appropriate signal processing. The separation of these orthogonal waveforms from $N_T$ transmitters require the suppression of autocorrelation sidelobes for each specific waveform as well as the corresponding cross-correlation sidelobes due to other $N_T-1$ waveforms.\\
	The orthogonal waveforms for MIMO radars are commonly generated with the following slow time approaches: (1) Time Division Multiplexing (TDM) where multiple transmitters are sequential / time staggered turned ON for consecutive sweeps up to the number of transmitters to achieve orthogonality (multiple transmitters are not turned ON simultaneously), resulting in reduced SNR and (2) Code Division Multiplexing (CDM) where multiple transmitters are turned ON simultaneously using outer phase coding (Hadamard matrix) across pulses to achieve orthogonality thereby, higher SNR is achieved. The detailed comparison of different orthogonal waveforms for a MIMO radar have been comprehensively covered in \cite{Wit.orthogonal,Xue.2009,Sun.H,Rabid}.\\ 
	Stretch processing based radars are commonly employed in various fields of operation such as remote sensing (SAR / ISAR), automotive, missile seekers etc. due to the reason that it reduces the computational bandwidth on receive while providing the same advantage of wide bandwidth radars limited only by the range interval associated with the reduced bandwidth of the receiver. For a stretch processing receiver, the matched filter is the Discrete Fourier Transform (DFT) of the received pulse as the range interval window is the associated set of beat frequencies within the intermediate frequency (IF) of the receiver. The target peaks after the Fast Fourier Transform (FFT - efficient implementation of DFT) are sinc patterned peaks which have high sidelobes (PSL at -13.2dB) associated with them and can result in masking of weaker targets. \\
	Traditionally, windowing techniques are usually employed to reduce these sidelobes but, at the cost of increased mainlobe width and loss in Signal-to-Noise-Ratio (SNR). Although, very good orthogonality can be achieved between MIMO waveforms using either TDM or CDM, yet even small residual non-orthogonalities between waveforms caused due to slight mismatch result in improper decoding of the received pulses which subsequently increase the interference level associated with the sidelobes. The traditional windowing techniques are robust to a certain degree of mismatch caused due to doppler shifts but, as shown in this paper, they also give degraded performance for ST-PC MIMO radar under other phase distortions due to the spatial location of targets \cite{Rabid,Blunt.2016}. \\
	To address this, mismatch filter (MMF) design \cite{Li.2008,Hua.2013,Ma.2010,hu2010optimal,stoica2008binary} usually coupled with waveform design is used to provide additional sidelobe suppression for MIMO waveforms separation at the receiver. In MMF, the received pulses are filtered using a modified matched filter with optimum designed weights instead of the traditional matched filter (unit weight vector) \cite{ghani1973}. It is pertinent to highlight that this mismatched based filtering is not a substitute for phase history-based windowing rather these MMFs are employed to provide additional sidelobe mitigation while applied in conjunction with traditional windowing techniques.\\
	We consider the design of optimum filter weights for the received pulses to mitigate the interference associated with ST-PC FMCW MIMO waveforms thereby, providing separation of the transmitted waveforms at the receiver with improved Signal-to-Interference-plus-Noise Ratio (SINR). We also consider the effect of doppler and other phase distortions caused due to improper decoding of the received pulses as, it is not possible to design waveforms which are orthogonal at all delays, dopplers and spatial angles \cite{Rabid}.\\ 
	Several sidelobe reduction techniques by designing optimized mismatched filters and weight functions have been used in the past for both conventional \cite{ghani1973,Blunt.APC,Harnett.SP,Kajenski.2016,Sun.2019} as well as MIMO radars \cite{Hua.2013,Ma.2010,hu2010optimal,Aittomaki.2017,Kim.2019,R.Feger.2016}. The traditional mismatched filter design for correlation receiver involves the use of a convolution matrix (designed using the replica of reference waveform) for the phase coded radar signals which is slightly different from the design of mismatched filter for stretch processing receiver which involves the use of DFT matrix (Matched filter bank) with FM waveforms as the radar signals.  The correlation-based MIMO radar techniques usually design mismatched filter formulations using optimization-based algorithms such as convex optimization \cite{boyd2004convex} without considering the doppler shifts of the target. Similarly, \cite{hu2010optimal,stoica2008binary} proposed a convex optimization-based filter design for sidelobe reduction in fast-time phase coded MIMO radar where no phase mismatch due to doppler shift was considered. Whereas on the other hand, with slow-time coding, doppler shifts associated with moving targets cannot be ignored as even a small mismatch due to doppler results in increased sidelobes. In the same context, \cite{Aittomaki.2017} considered the optimum filter design for nonzero doppler while using the convex optimization technique for ST-PC MIMO radar.\\
	Recently, the filter design techniques for stretch processing receivers have gained importance due to the availability of commercial stretch processing hardware for MIMO radars. In \cite{R.Feger.2016}, a weight window design method for sidelobes reduction using the convex optimization approach for Fast-Time Phase Coded (FT-PC) FMCW MIMO radar was proposed. A new approach to address the sidelobes reduction problem was covered by \cite{Harnett.SP} using an optimal Least Squares MMF which proposed the design of a compensated matrix in place of traditional DFT for a conventional stretch processing receiver. This paper also proposes an alternate adaptation of an Adaptive Pulse Compression (APC) algorithm \cite{Blunt.APC,Higgins.GCAPC} based on Minimum Mean Squared Error (MMSE) filter design which iteratively reduced the sidelobes to the noise floor using only a single range samples snapshot. We consider designing the compensation matrix based on \cite{Harnett.SP} and \cite{Hemmingsen.WD} for the ST-PC MIMO radar system where this compensation matrix is used to design optimal adaptive mismatch filter weights using the proposed algorithm \cite{Blunt.APC,Higgins.GCAPC,Blunt.RISR}.\\ 
	The proposed filter design method in this paper introduces a new method to achieve optimal waveform separation at the MIMO receiver using the concept of baseline Reiterative MMSE (RMMSE) APC algorithm \cite{Blunt.APC} and its multistatic versions \cite{McCormick.shared}. It is worth mentioning that the APC algorithm and its over the years improved versions only considered the sidelobes reduction for conventional radar. To date, to the best knowledge of the author, no mismatch filter design including the baseline RMMSE APC and its improved versions have been proposed for autocorrelation and cross-correlation sidelobes reduction for subsequent waveform separation on a ST-PC FMCW MIMO radar.\\  
	The baseline RMMSE APC algorithm \cite{Blunt.APC} provides closed-form expression resembling the MVDR expression. Unlike MVDR, it requires a single snapshot to calculate the power estimates of the estimated range profile in the previous iteration which is then used to design the filter weights for the current iteration. Within 1-4 iterations, the filter weights are optimized such that all the sidelobes other than the main lobe are suppressed to the noise floor. However, it has been demonstrated in the past that slight doppler mismatch along fast time samples can degrade the performance of sidelobes suppression. It can be inferred and later demonstrated in this paper, that any other phase distortion such as the spatial location of targets, improper waveform decoding etc. would also degrade the performance of the baseline RMMSE filter. We have derived the mathematical expressions for the possible phase distortions that need to be compensated for subsequent optimal implementation of APC filter. Moreover, closed-form mathematical expression has also been derived for the proposed APC algorithm for MIMO radar. \\
	ST-PC FMCW MIMO radars require doppler compensation for moving targets for proper decoding of received phase coded pulses for subsequent accurate angle estimation of targets \cite{Kim.2019,Gonzalez.,TexasInstruments.,Bechter.motion}. The use of the APC algorithm on received pulses through traditional order of operations always provide degraded performance as the received pulses  contain random phase and doppler contributions. This paper proposes to re-adjust the order of operations in a ST-PC FMCW MIMO radar by implementing the proposed RMMSE APC algorithm on coherently integrated angle-doppler range samples thereby, providing optimal sidelobes reduction. Consequently, greater reduction in sidelobes can be achieved due to sidelobes decoherence as the doppler and angle processed pulses are integrated while, additional fidelity for target peaks is also achieved.\\ 
	This paper is organized as follows. Section II covers the signal model and its derivation for ST-PC FMCW MIMO radar waveforms. It also provides details related to the decoding procedure and proposed radar signal processing chain to set the initial cost function for optimal range sidelobes minimization. In Section-III, the APC based on RMMSE filter design for considered MIMO radar signal model is derived and modelled. Section-IV covers the simulated response for the APC based on MMSE filter design for MIMO radar which is followed by the recommendation of the best methodology to achieve optimal sidelobe suppression for MIMO radars using APC. Section-V then discusses the optimal methodology for sidelobes reduction for MIMO radars while examining their effects both in simulated and real measured data. Finally, Section-VI concludes the paper. 
	
	\section{MIMO Radar Signal Model for Stretch Processing Receiver}
	\subsection{FMCW Waveform Model}
	The FMCW waveform generation over a single frame transmitted from each of the antennas is the  train of FMCW waveforms over which appropriate fast / slow time MIMO coding is carried out to achieve orthogonality. General form of single FMCW waveform, defined over the interval ${0}\leq$$t\leq{T_p}$ for chirp duration $T_p$ is expressed as
	\begin{equation}u(t)=\exp \left(j\left(2 \pi f_{\mathrm{c}} t+\pi K t^{2}\right)\right)
		\label{lfm}
	\end{equation}
	\begin{equation}s(t)=\sum_{m=-\infty}^{\infty} u\left(t-m T_{\text {p }}\right)
		\label{lfmtrain}
	\end{equation}
	where $f_c$ is the ramp starting frequency and $K=B / T_{\text {p }}$ being the chirp rate. The returned signal $y(t)$ after illumination from the range profile and reflected back from a moving target at a delay $\tau$, doppler frequency $f_d$ and angle of arrival $\theta$ (with reference to transmit / receive array geometry center)  in the presence of Additive White Gaussian Noise (AWGN) $\gamma(t)$ is processed as a collection of $N_p$ pulses within a Coherent Pulse Interval (CPI) where each pulse is dechirped as it passes through the mixer stage subsequently followed by bandpass filtering and IQ demodulation at the each receiver. Consequently, the dechirped signal for $m^{th}$ pulse and $n^{th}$ receiver is the combination of sinusoidals at different frequencies (sum of beat and doppler frequency of target(s)). This dechirped signal can be expressed as
	\begin{equation}\begin{aligned}
			y^{m,n}(t)&= \alpha(t)\exp \left(j 2 \pi\left(f_{B} t+m f_{d} T_{\mathrm{p}}+\frac{ud_{R}n}{\lambda} \right)\right)\\ &+\gamma^{m,n}(t)
	\end{aligned}\end{equation}
	where $\alpha(t)$ is the target scattering amplitude, $f_B$ is the target beat frequency related to target two way propagation delay $\tau$ and doppler frequency $f_d$ by $f_B=K\tau+f_d$, $u$ is the target angle of arrival given by $sin(\theta)$, $\lambda $ is the radar wavelength, $d_R$ is the spacing between the receivers and $\gamma^{m,n}(t)$ being the dechirped noise for $m^{th}$ pulse at the $n^{th}$ receiver. The dechirped signal is then converted to digital domain using Analog to Digital Converter (ADC) with sampling frequency $f_s=1/T_s$. By considering $N_f=T_p/T_s$ samples per pulse per receiver for a total of $N_p$ pulses and $N_r$ receivers in a single frame, the digital samples representation of dechirped signal is as follows
	\begin{equation}
		\begin{aligned}
			y(q,m,n)&=\alpha(t) \exp \left(j 2 \pi\left(\hat{f}_{B} q+\hat{f}_{d} m+\hat{f}_{\theta}n\right)\right)\\&+\gamma(q, m,n)
		\end{aligned}
		\label{sisoreceived}
	\end{equation}
	where the discretized versions along each dimension is given by $q=0, \cdots, N_f-1$,  $m=0,1, \cdots, N_{p}-1$, $n=0,1, \cdots, N_{r}-1$. Consequently, the sampled version of normalized beat, doppler and spatial frequency is given by $\hat{f}_{B}=f_{B} T_{s}$, $\hat{f}_{d}=f_{d}  T_{\text {p }}$ and $\hat{f}_{\theta}=\frac{ud_{R}}{\lambda}$, respectively.\\  
	Following the traditional sequence of operations, the bank of matched filters tuned to different delays, doppler shifts, and spatial frequencies along each dimension is typically applied to obtain the range, doppler, and angle estimates of the target(s). For a stretch processing receiver, Discrete Fourier Transform (DFT - implemented using Fast Fourier Transform (FFT)) is the matched filter and results in the superposition of sinc peaks at the detected target beat frequencies. Afterwards, an FFT along the slow time dimension is used to detect the doppler shifts associated with moving target(s). Beamforming is performed to estimate the angle of arrival of the target(s) by applying FFT along the spatial domain.
	
	\subsection{Slow Time Phase Coded MIMO Radar Signal Model}
	For the considered MIMO radar, let's consider $N_T$ transmitters and $N_R$ receivers. FMCW waveform defined in \eqref{lfmtrain} is employed as the common waveform $s(t)$ from all transmitters. To implement orthogonal waveforms in slow time domain, the initial phase for each pulse is applied as per the outer Hadamard coding matrix \eqref{hadamard} for each transmitter. Keeping in view the hardware implementation, only Binary Phase Modulation is considered which can be easily implemented on commercially available hardware. The outer Hadamard matrix $A$ of any order $k$ is a $k\times k$ matrix populated with 1s and -1s while maintaining the property $AA^H=kI_k$ where $I_k$ is $k \times k$ identity matrix. Any two rows of such a Hadamard matrix are orthogonal to each other and the corresponding Walsh-Hadamard codes of each row can be used to modulate the repeating common waveform pulses such as FMCW from different transmitters to achieve orthogonality in the slow time domain.\\
	\begin{equation}A_{2}=\left[\begin{array}{cc}
			1 & 1  \\
			1 & -1 
		\end{array}\right]
		\label{hadamard}\end{equation}
	Since the orthogonality is achieved by modulation of binary phases on the basic FMCW waveform so the ambiguity function properties of the FMCW signal are retained while providing the additional benefits associated with the angular diversity of MIMO radar. It is pertinent to mention that the chirps are generated using a common reference source (single ramp generator) which ensures phase coherence among multiple transmissions. Due to BPM, the signal model incorporating orthogonal FMCW waveform transmission is remodelled using \eqref{hadamard} and expressed as
	\begin{equation}s(t)=\sum_{m=0}^{N_{p}-1} \sum_{i=0}^{N_{T}-1} u(t-m T_p) \exp (j \phi_m(i))\end{equation}
	where $\phi(m, i) \rightarrow (0,\pi)$ implemented according to the generated hadamard matrix $A_k$ of order $k$. Considering $Z$ point scatterers, the discretized received signal can be expressed using \eqref{sisoreceived} as follows
	\begin{equation}
		\begin{aligned}
			y(q, m, n) &=\sum_{z=1}^{Z}\alpha_z \sum_{i=0}^{N_{T}-1} \exp (j 2 \pi(\hat{f}_{B} q+\hat{f}_{d} m\\
			& + \frac{(i) d_{T}+(n)d_{R}}{\lambda}u_z))  \exp(j\phi_m(i))\\&+\gamma(q, m, n) \end{aligned}
		\label{mimoreceived}
	\end{equation}
	where $d_{T}=N_{R} \lambda / 2$ and $d_{R}=\lambda / 2$. If we assume doppler shift $f_d$ and angle of arrival $\theta$ of a single point scatterer to be zero i.e. static target at the center of array geometry at a certain range $R$ related to delay $\tau$ by $\tau=2R/c$ then \eqref{mimoreceived} for received MIMO signal reduces to
	\begin{equation}
		\begin{aligned}
			y(q, m, n)&=\alpha_1 \sum_{i=0}^{N_{T}-1} \exp (j  (2\pi\hat{f}_{B} q+\phi_m(i))\\&+\gamma(q, m, n) 
		\end{aligned}
		\label{mimoreceivedonlypc}
	\end{equation}
	which is a complex sinusoid related to beat frequency (lying within the designed range interval - IF bandwidth) of target further scaled by the scatterer's amplitude and phase. Moreover, it also contains additional phase shifts due to simultaneous orthogonal binary phase modulated target echoes from multiple transmitters. 
	
	\subsection{Matched Filter Bank Formulation using Compensation Matrix}
	The normalized matched filter for the captured receive samples of length $N_f$ at the $n^{th}$ receiver for each respective delay i.e. $l=0,1,\cdots,L-1$ within the designed range swath i.e. [$R_{near}$ $R_{far}$] can be discretized similar to \cite{Harnett.SP} to form the compensation matrix $\mathbf{F}$. The $N_f \times L$ compensation matrix $\mathbf{F}$ is used as a bank of matched filters in place of traditional DFT to design filter weights with adaptive pulse compression while providing enhanced range sidelobes suppression. Here it is assumed oversampling of range interval by a factor of $K$ relative to 3 dB bandwidth of the FMCW waveform to provide additional fidelity subsequently required for adaptive mismatched filter formulation. This oversampling relation can be expressed as 
	\begin{equation}
		\begin{aligned}
			L=KN_f\\
			N_f=T_p/T_s
		\end{aligned}
	\end{equation}
	The expression for $\mathbf{F}$ with oversampled range swath is as follows
	{\small \begin{equation}\mathbf{F}=[\mathbf{w}(R_{near}) \quad \mathbf{w}(R_{near}+\delta R) \cdots \mathbf{w}(R_{near}+(L-1) \delta R)]\end{equation}
	}
	where $\mathbf{w}(R_{near}+l(\delta R))$ is a $N_f \times 1$ vector and is computed for each delay $l$ which subsequently forms the columns of $\mathbf{F}$ using
	\begin{equation}
		\mathbf{w}(R_{near}+l(\delta R))=\exp(-j2{\pi}(f_{near}+l(\delta f))\mathbf{t})
	\end{equation}
	where $f_{near}$ corresponds to the beat frequency related to $R_{near}$ by $K(2R/c))$, $\delta f$ is the corresponding frequency sample spacing and $\mathbf{t}$ is $N_f \times 1$ vector given by 
	\begin{equation}
		\mathbf{t}=[\tau_{near}\quad \tau_{near}+T_s\quad \cdots\quad \tau_{near}+(N_f-1)T_s]
	\end{equation}
	where $\tau_{near}$ corresponds to beginning of range interval i.e. $R_{near}$. The resultant compensation matrix corresponding to range interval $(0\sim R_{max})$ can be derived as
	\begin{equation}\resizebox{.88\hsize}{!}{$
			\mathbf{F}=\left[\begin{array}{cccc}
				W^{(0)(0)}&W^{(0)(1)}&\cdots&W^{(0)(L-1)}\\
				W^{(1)(0)}&W^{(1)(1)}&\cdots&W^{(1)(L-1)}\\
				\vdots&\vdots&\ddots&\vdots\\
				W^{(N_f-1)(0)}&W^{(N_f-1)(1)}&\cdots&W^{(N_f-1)(L-1)}
			\end{array}\right]$} 
	\end{equation}\normalsize
	\begin{equation}{W^{(q)(l)}}=\frac{1}{\|{\mathbf{w}{(l)}}\|_{2}} {{w}^{(q)(l)}}\end{equation}
	where  ${W^{(q)(l)}}$ represents the compensated sample $q$ at delay $l$. For the received samples of $m^{th}$ pulse from $N_t$ transmitters at the $n^{th}$ receiver, the \eqref{mimoreceivedonlypc} can be modelled to incorporate the compensation matched filter bank formulation by
	\begin{equation}
		\begin{aligned}
			\mathbf{y}^{m,n}&= \sum_{i=0}^{N_{T}-1} \mathbf{F} \mathbf{x}_{i}\exp (j \phi_m(i))+\mathbf{\gamma}^{m,n}
			\label{mimocomp_pc}\end{aligned}
	\end{equation}
	where $\mathbf{y}^{m,n}$ is $N_f \times 1$ received vector, $\mathbf{x_i}$ is the oversampled length $L$ complex scattering coefficients over range swath for the $i^{th}$ transmitter. The received signal model for a single CPI at the $n^{th}$ can then be expressed as
	\begin{equation}
		\begin{aligned}
			\mathbf{Y}^{n}& =\sum_{i=0}^{N_{T}-1}[\mathbf{y}_{0}^{n}\exp (j \phi_0(i)) \quad \mathbf{y}_{1}^{n}\exp (j \phi_1(i))\\ 
			&\cdots \mathbf{y}_{N_p-1}^{n}\exp (j \phi_{N_p-1}(i))]
		\end{aligned}
		\label{wodopplerrd}
	\end{equation}
	This received data matrix $\mathbf{Y}^{n}$ is then processed to obtain range-doppler data followed by decoding for transmitted waveform separation to perform improved angle estimation due to MIMO waveforms. For a single moving target at delay $\ell$ with doppler frequency $\hat{f}_d$, \eqref{mimocomp_pc} can be expressed as
	\begin{equation}
		\begin{aligned}
			\mathbf{y}^{m,n}(\ell,\hat{f}_{d})&=
			\sum_{i=0}^{N_{T}-1} \mathbf{F} \mathbf{x}_{i}\exp (j (2\pi \hat{f}_{d} (m))+ \phi_m(i))\\&+\mathbf{\gamma}^{m,n}(\ell)
		\end{aligned}
	\end{equation}
	Like \eqref{wodopplerrd}, this can be extended for collection of $N_P$ echoes within a CPI yielding the following matrix
	\begin{equation}{
			\begin{aligned}
				\mathbf{Y}^{n}(\ell, \hat{f}_{d})&=\sum_{i=0}^{N_{T}-1}[\mathbf{y}_{0}^{n}(\ell)e^{j \phi_0(i)} \quad \mathbf{y}_{1}^{n}(\ell)e^{j (2\pi \hat{f}_{d} + \phi_1(i)})\\ 
				&   \cdots \mathbf{y}_{N_p-1}^{n}(\ell)e^{j (2\pi \hat{f}_{d} (N_p-1)+ \phi_{N_p-1}(i))}]
		\end{aligned}}\label{mimoreceived_doppler}
	\end{equation}
	
	\subsection{Decoding of Slow-Time Coded Received Waveforms}
	The additional superposition of phase shifts in \eqref{mimoreceived_doppler} from all transmitters in a slow-time coded MIMO radar need to be decoded for accurate angular estimation with benefit of enhanced angular resolution. These phase shifts can be decoded by considering the captured number of pulses within the transmit block duration ($\delta t=N_T \times T_P$) when the chirp duration is identical across all transmitters. As the code is repeated after each block duration, therefore, the maximum ambiguous doppler is subsequently reduced by $N_T$ times as compared to a SIMO radar \cite{Sun.H}. \\ For a simple case of two transmitters with a single static target at delay $\ell$ and angle $\theta$, the captured echoes within block duration consist of two consecutive pulses i.e. $\mathbf{s_{A}}$ and $\mathbf{s_{B}}$. Based on hadamard coding matrix, it can be observed that each of these received pulses within first block duration has the following form at the $n^{th}$ receiver
	\begin{equation}
		\begin{aligned}
			\mathbf{s_{A}}^{n}(\ell)&=\sum_{i=0}^{N_{T}-1}\mathbf{y}_{0}^{n}(\ell)e^{j \phi_0(i)}\\&=(\mathbf{s_{1}}+\mathbf{s_{2}}e^{j2\pi(\frac{ d_Tu_1}{\lambda})})e^{j2\pi(\frac{ n d_R u_1}{\lambda})} \\
			\mathbf{s_{B}}^{n}(\ell)&=\sum_{i=0}^{N_{T}-1}\mathbf{y}_{1}^{n}(\ell)e^{j \phi_1(i)}\\&=(\mathbf{s_{1}}-\mathbf{s_{2}}e^{j2\pi(\frac{ d_Tu_1}{\lambda})})e^{j2\pi(\frac{ n d_R u_1}{\lambda})}
			\label{mimo_stpctx2}\end{aligned}
	\end{equation}
	The transmissions from two transmitters i.e. $\mathbf{s_{1}}$ and $\mathbf{s_{2}}$ need to be separated to exploit the virtual array structure of MIMO radar. Through algebraic simplifications, these transmissions can be easily separated to give the following expressions
	\begin{equation}\begin{array}{l}
			\mathbf{s_{1}}=\dfrac{\mathbf{s_{A}}^{n}(\ell)+\mathbf{s_{B}}^{n}(\ell)}{2} \\
			\hat{\mathbf{s_{2}}}=\dfrac{\mathbf{s_{A}}^{n}(\ell)-\mathbf{s_{B}}^{n}(\ell)}{2}
		\end{array}
		\label{decodingeqn}\end{equation}
	where $\hat{\mathbf{s_{2}}}=\mathbf{s_2}\exp(j2\pi(\frac{ d_Tu_1}{\lambda})))$. 
	For a moving target scenario, \eqref{mimo_stpctx2} contains an additional phase shift associated with the second captured pulse i.e. $\mathbf{s_B}$ and the resultant expressions can be written as
	\begin{equation}\begin{array}{l}\resizebox{.6\hsize}{!}{$
				\mathbf{s_{A}}^{n}(\ell, \hat{f}_{d})=(\mathbf{s_{1}}+\mathbf{s_{2}}\exp(j2\pi(\frac{ d_Tu_1}{\lambda})))$} \\
			\resizebox{0.8\hsize}{!}{$\mathbf{s_{B}}^{n}(\ell, \hat{f}_{d})=(\mathbf{s_{1}}-\mathbf{s_{2}}\exp(j2\pi(\frac{ d_Tu_1}{\lambda})))\exp (j (2\pi \hat{f}_{d}))$}
			\label{capturedwithdoppler} \end{array}\end{equation}\normalsize
	Now, the decoding for separation of transmissions can not be applied through algebraic simplifications and rather require the compensation of doppler shift associated with the second captured pulse. Only then, proper decoding can be implemented thereafter giving accurate angle estimation. To cater for this, traditional slow-time coded MIMO radars usually employ separate 2D range-doppler processing with block duration ($\delta t$) as the Pulse Repetition Interval (PRI) for each repeating coded sequences ($\mathbf{s_{A}}$, $\mathbf{s_{B}}$) within a CPI at the receiver. This is followed by an appropriate threshold detector to extract the range and doppler bins of the detected target(s) for subsequent doppler compensation.  The general expression for doppler compensation for $N_T$ transmitters is follows
	\begin{equation}\psi_i(\hat{f_d})={-2 \pi i\hat{f_d}}, \quad  i = 0,1 \cdots N_T-1\end{equation}
	where $i$ corresponds to consecutive received pulses within block duration ($\delta t$) i.e. $0 \rightarrow{A}$, $1 \rightarrow{B}$ and so on. 
	\subsection{Adjustment of Order of Operations and Phase Distortions Compensation for Mismatch Filter Formulation}\label{oop}
	From literature, it is observed that adaptive pulse compression using Minimum Squared Error structure gives degraded performance in the presence of phase distortions i.e. doppler \cite{Cuprak.2017}. Similarly for MIMO radars, the works in \cite{Rabid,Blunt.2016} highlight the role of the spatial location of targets which also adds to the additional phase distortion that also needs compensation for proper decoding of the target returns. The degradation due to spatial location of targets was analyzed in \cite{davis2015minimum,goodman2015angle} where it was recommended to consider the target angles for optimum suppression of range sidelobes in MIMO radars. The combined effect of these phase distortions considerably degrades the performance of the RMMSE filter as shown in the next section. Therefore, implementing APC based filter on \eqref{capturedwithdoppler} does not provide optimal range sidelobes reduction. To address this, these phase distortions need to be compensated before the implementation of APC based filter for optimal range sidelobes reduction. Whereas, the phase shift due to doppler and angle can only be compensated after the application of DFT along the slow-time and receive array dimensions.\\
	To overcome this, this paper proposes to re-adjust the traditional order of operations for carrying out 3D-FFT along range, doppler, and angle domain where FFT being linear operator can be re-adjusted while having no effect on the traditional radar signal processing. This re-adjustment of order of operations provide improved dynamic range where additional sidelobes suppression is achieved due to sidelobes decoherence effect. This sidelobes decoherence have also been utilized previously in \cite{jakabosky2016spectral} for mitigation of range sidelobes. Moreover, the additional fidelity due to coherent integration of doppler processed pulses was previously utilized in \cite{McCormick.shared}. According to the proposed arrangement, the raw received data within a CPI at the $n^{th}$ receiver is the first doppler processed to get the   $N_f \times N_P$ pulse-doppler 2D matrix where each column corresponds to the discretized doppler shifts. The captured echoes $\mathbf{s_{A}}^{n}(\ell)$ and $\mathbf{s_{B}}^{n}(\ell)$ can be extended to incorporate all the pulses within a CPI and expressed as
	\begin{equation}
		\begin{aligned}
			\mathbf{S_A}^{n}(\ell, \hat{f}_{d})&=[\mathbf{s}_{A}^{n}(\ell) \quad \mathbf{s}_{A}^{n}(\ell)e^{j(2\pi \hat{f}_{d}(2))}\\ 
			&   \cdots \mathbf{s}_{A}^{n}(\ell) e^{j(2\pi \hat{f}_{d} 2(N_p-1))}]\\
			\mathbf{S_B}^{n}(\ell, \hat{f}_{d})&=[\mathbf{s}_{B}^{n}(\ell)e^{j (2\pi \hat{f}_{d})}\quad \mathbf{s}_{B}^{n}(\ell)e^{j (2\pi \hat{f}_{d}(3))}\\ 
			&   \cdots \mathbf{s}_{B}^{n}(\ell)e^{j (2\pi \hat{f}_{d} (2N_p-1))}]
		\end{aligned}
	\end{equation}\normalsize
	The coherent received data vector after the implementation of DFT for the discretized doppler  $\hat{f_d}$ is expressed as 
	\begin{equation}\mathbf{S_A}(\ell, \hat{f_d})=\frac{1}{N_P} \mathbf{S_A}(\ell, \hat{f_d})\left(\mathbf{h} \odot \mathbf{a}\left(N_T\hat{f_d}, \lambda\right)\right)\end{equation}
	where $N_p \times 1 $ vector $\mathbf{h}$ represents the window across the doppler domain to reduce the corresponding doppler sidelobes,  $\odot$ represents the elementwise hadamard product and the $N_p \times 1 $ vector $\mathbf{a}$ is the corresponding FFT for the doppler shift $N_T\hat{f_d}$ given by
	\begin{equation}\resizebox{.88\hsize}{!}{$\mathbf{a}\left(N_T\hat{f_d}, \lambda\right)=\left[\begin{array}{ccc}
				1 & e^{-j {2 \pi N_T\hat{f_d}} }&\left.\cdots e^{-j(N_P-1) {2 \pi N_T\hat{f_d}}}\right]^{T}
			\end{array}\right.$}\end{equation}\normalsize
	The coherent received $N_f \times N_P$ pulse-doppler matrix $\mathbf{S_{Adop}}$ for the full discretized doppler domain can be obtained by implementing the corresponding bank of DFT filters for $\mathbf{f_{dop}}=[-\hat{f_d}(max) \cdots \hat{f_d}(max)]$. Similar procedure is applied to obtain the other 2D pulse-doppler matrix ($\mathbf{S_{Bdop}}$) during a single CPI.\\
	This is followed by threshold detection (e.g. Cell Averaging Constant False Alarm Rate) of the doppler processed 2D matrix to find the doppler bins of the detected targets. Afterwards, corresponding doppler compensation is applied  to each detected doppler bin on the 2D pulse doppler matrix ($\mathbf{S_{Bdop}}$), so that, decoding may be applied afterwards using \eqref{decodingeqn}.  As a result, the corresponding received doppler processed 2D pulse-doppler matrices for each transmitter i.e. $\mathbf{S_{1}}, \mathbf{S_{2}}$ are successfully separated.\\
	\begin{equation}\begin{array}{l}
			\mathbf{S_{1}}=\dfrac{\mathbf{S_{Adop}}+\mathbf{S_{Bdop}}}{2} \\
			\hat{\mathbf{S_{2}}}=\dfrac{\mathbf{S_{Adop}}-\mathbf{S_{Bdop}}}{2}
		\end{array}
		\label{compensatedtxns}\end{equation}
	Followed by doppler processing and subsequent phase compensation, angle processing can be performed by stacking the decoded received 2D pulse-doppler matrices corresponding to each transmitter resulting in a $N_{N_T \times N_R} \times N_f \times N_P$ 3D data cube. Traditional DFT based angle processing is performed followed by coherent integration across doppler and angular domains. Finally, the 1D coherently summed fast-time samples are taken as input to the proposed RMMSE base adaptive pulse compression filter for robust mitigation of range sidelobes.
	\section{Mismatch Filter Formulation for Range Sidelobes Reduction in MIMO Radars}
	The mismatch and slight distortions during phase compensation and decoding may result in imperfect waveform separation at the receiver which results in increased range sidelobes. To cater for this, Minimum Mean Squared Error (MMSE) based filter was modified to incorporate these mismatches induced due to transmissions from other MIMO waveforms. Consequently, the modified RMMSE based APC filter for MIMO radar waveforms was derived based on the proposed order of operation in the previous section.\\
	The mismatched filter formulation employed for MIMO radar in this paper is based on gain constrained Minimum Mean Squared Error (MMSE) optimization approach \cite{Higgins.GCAPC}. The Reiterative MMSE (RMMSE) approach was modelled to incorporate the slow-time coded MIMO radar. RMMSE filter uses a single snapshot of range samples to reduce the cross-correlation and autocorrelation sidelobes to noise floor in an iterative manner thereby implementing the pulse compression adaptively. As discussed in previous section, the coherently integrated decoded range samples obtained after doppler and angle processing defined in \eqref{compensatedtxns} would be used as the initial cost function for MMSE filter optimization for a ST-PC MIMO radar.
	\subsection{Derivation of RMMSE Filter for MIMO radar}
	The matched filter outputs, for each decoded transmission $\mathbf{s}_i$ (within block duration $\delta t$) separated from received pulses within a CPI across all the receivers,  after the coherent integration of  angle-doppler processed response is obtained by
	\begin{equation}
		\hat{\mathbf{x}}_{i,MF}=\mathbf{F}^{H} \mathbf{s}_i, \qquad i=0,1 \cdots N_T-1
		\label{mf}\end{equation}
	The matched filter response obtained above is simply an oversampled estimate of complex scattering coefficients for the defined range swath in comparison to the response obtained through FFT implementation with $(\mathbf{\bullet})^H$ being the Hermitian (complex-conjugate) operator. An adaptive MMSE filter bank  \cite{haykin2005adaptive} can be obtained for the signal model in \eqref{mf} by  using Reiterative Super Resolution Algorithm (RISR) \cite{Blunt.RISR}. The cost function based on gain-constrained MMSE can be derived as
	\begin{equation}\resizebox{.88\hsize}{!}{$\begin{array}{ll}
				\underset{\mathbf{F}^*}{\operatorname{minimize}} & E\left[\left|\mathbf{x}-\mathbf{F}^{H} \mathbf{s_i}\right|^2\right] \\
				\text { subject to } & \mathbf{f_{MIMO,i}(\ell)}^{H} \mathbf{f(\ell)}=1, \: \ell = 0,1 \cdots L-1
			\end{array}$}\end{equation}\normalsize
	The cost function with the unity gain constraint can be expressed as
	\small\begin{equation}J=E\left[\left|\mathbf{x}-\mathbf{F}^{H} \mathbf{s_i}\right|^2\right]+Re \left\{\Lambda(\mathbf{f_{MIMO,i}(\ell) ^{H} \mathbf{f(\ell)}-1})\right\}\end{equation}\normalsize
	where $\mathbf{f(\ell)}$ is the $\ell^{th}$ column of $N_f \times L$ matrix $\mathbf{F}$, $\Lambda$ is the Lagrange multiplier, $Re(\mathbf{\bullet})$ is the real operator and  $E(\mathbf{\bullet})$ is the expectation. The general solution with minimized cost function is given by
	\begin{equation}\mathbf{F}_{\mathrm{MIMO,i}}=\left(E\left\{\mathbf{s_i} \mathbf{s_i}^{H}\right\}\right)^{-1} E\left\{\mathbf{s_i} \mathbf{x}^{H}-\frac{\Lambda}{2}\mathbf{F}\right\} 
		\label{risr}\end{equation}
	where each decoded transmission $\mathbf{s}_i$ can be approximately represented as
	\begin{equation}\mathbf{s}_i=\mathbf{F} \mathbf{x}_{i}+\gamma
		\label{receivedpulse}\end{equation}
	where $\mathbf{\gamma}$ is the $N_f \times 1$ noise vector. The substitution of \eqref{receivedpulse} in \eqref{risr} under the assumption of  statistically uncorrelated scatterers provides the following solution to the expectations which subsequently results in a Minimum Power Distortion Response closed-form solution 
	\begin{equation}E\left[\mathbf{s_i} \mathbf{x}_{i}^{*}\right]=E\left[\left|\mathbf{x}_{i}\right|^{2}\right] \mathbf{F} =\left|\mathbf{x}_{i}\right|^{2} \mathbf{F}\end{equation}
	\begin{equation}
		\resizebox{.88\hsize}{!}{$E\left[\mathbf{x}_{i} \mathbf{x}_{j}^{*}\right]=\left\{\begin{array}{c}\begin{aligned}
					&E\left[\left|\mathbf{x}_{i}\right|^{2}\right] \text { for } i=j, \quad \{i,j\} = 0,1 \dots N_T-1\\
					&\qquad 0 \qquad \text { otherwise }
			\end{aligned}\end{array}\right.$}\end{equation}\normalsize
	\begin{equation}\mathbf{f}_{\mathrm{MIMO, i}}(\ell)=\frac{\left(\mathbf{F} \mathbf{P}_{i} \mathbf{F}^{H}+\mathbf{R}\right)^{-1} \mathbf{f}_{\ell}}{\mathbf{f}_{\ell}^{H}\left(\mathbf{F} \mathbf{P}_{i} \mathbf{F}^{H}+\mathbf{R}\right)^{-1} \mathbf{f}_{\ell} }\end{equation}
	where
	\begin{equation}\begin{aligned}
			\mathbf{R} &=\mathbf{F}\left(E\left[\mathbf{x}_{i, k-1} \mathbf{x}_{i, k-1}^{H}\right]\right) \mathbf{F}^{H}+E\left[\mathbf{u} \mathbf{u}^{H}\right] \\
			&=\mathbf{F} \mathbf{P}_{i, k-1} \mathbf{F}^{H}+\sigma_{\gamma}^{2} \mathbf{I}_{N_{f} \times N_{f}}
	\end{aligned}\end{equation}
	and
	\begin{equation}\mathbf{P}_i=E\left[\mathbf{x}_i \mathbf{x}_i^{H}\right]\end{equation}
	where $\sigma_{\gamma}^{2}$ is the noise power. If we include the covariance matrix estimates corresponding to the matched filter estimates of other transmitters, this would provide additional fidelity for the target peaks while also providing sidelobes decoherence effect which provide addtional sidelobes suppression \cite{McCormick.shared,jakabosky2016spectral}. The modified covariance matrix incorporating the power estimates from other transmitters is given by
	\begin{equation}\begin{aligned}
			\mathbf{R} &=\mathbf{F}\left(E\left[\mathbf{x}_{i,k-1} \mathbf{x}_{i,k-1}^{H}\right]\right) \mathbf{F}^{H}+E\left[\mathbf{u} \mathbf{u}^{H}\right]\\&+\sum_{j=0,i\neq j}^{N_T -1}\mathbf{F}\left(E\left[\mathbf{x}_{j,MF} \mathbf{x}_{j,MF}^{H}\right]\right) \mathbf{F}^{H} \\
			&=\mathbf{F} \mathbf{P}_{i,k-1} \mathbf{F}^{H}+\sigma_{\gamma}^{2} \mathbf{I}_{N_{f} \times N_{f}}+\sum_{j=0,i\neq j}^{N_T -1}\mathbf{F} \mathbf{P}_{j} \mathbf{F}^{H}
	\end{aligned}\end{equation}
	The range profile is not known \textit{a-priori} and it can be estimated through iterative estimation technique given as
	\begin{equation}\hat{\mathbf{P}}_{i,k}=\left(\hat{\mathbf{x}}_{\mathrm{MIMO}, k-1} \hat{\mathbf{x}}_{\mathrm{MIMO}, k-1}^{H}\right) \odot \mathbf{I}_{L \times L}\end{equation}
	where $\hat{\mathbf{P}}_{i,k}$ is the $k^{th}$ estimate of received range profile estimate related to $i^{th}$ transmitter while $\hat{\mathbf{x}}_{\mathrm{MIMO}, k-1}$ is the previous range profile estimate. The initial estimate i.e. $k=0$ is critical in reducing the number of iterations in achieving the desired mean squared error criteria. Here, like \cite{Blunt.APC,Harnett.SP}, the initial estimate is set to the matched filter range profile estimate obtained using \eqref{mf} for each decoded transmission. Finally, the adaptive estimation of $k^{th}$ iteration for a certain received range samples snapshot is given by
	\begin{equation}\resizebox{.88\hsize}{!}{$\tilde{x}_{\mathrm{MIMO},k, i}(\ell)=\mathbf{f}_{\mathrm{MIMO, i}}^H(\ell) \mathbf{s_i}=\frac{\left(\mathbf{F} \mathbf{P}_{i} \mathbf{F}^{H}+\mathbf{R}\right)^{-1} \mathbf{f}_{\ell}}{\mathbf{f}_{\ell}^{H}\left(\mathbf{F} \mathbf{P}_{i} \mathbf{F}^{H}+\mathbf{R}\right)^{-1} \mathbf{f}_{\ell}}$}\end{equation}\normalsize
	Here $\tilde{x}_{\mathrm{MIMO},k, i}(\ell)$ is the RMMSE based adaptive estimate of  coherently averaged angle-doppler processed range profile for $i^{th}$ transmitter $\mathbf{s_i}$ at delay $\ell$. When extended to include all delays ($l=0,1,2 \cdots L-1$), it provides APC filter for complete range swath.
	\begin{table}[h]
		\centering
		\caption{Radar and Waveform Design Parameters for Simulation}
		\begin{tabular}{|l|c|c|}
			\hline
			\textbf{Parameter} & \textbf{Variable} & \textbf{Value} \\ \hline
			Number of Transmitters & $N_T$ & 2 \\ \hline
			Number of Receivers & $N_R$ & 4 \\ \hline
			Start Frequency & $f_c$ & 77 GHz \\ \hline
			Waveform Bandwidth & $B$ & 240 MHz \\ \hline
			Sweep Time & $T_p$ & 2.67 $\mu$s \\ \hline
			Chirp Rate & $K$ & 90 MHz/$\mu$s \\ \hline
			ADC Sampling Frequency & $f_s$ & 80 MHz \\ \hline
			Number of Chirps in CPI & $N_p$ & 32 \\ \hline
			Near Range & $R_{near}$ & 0 \\ \hline
			Far Range & $R_{far}$ & 133 m \\ \hline
			Oversampling factor & $k$ & 3 \\ \hline
		\end{tabular}
		\label{tab:fmcw_radar}
	\end{table}
	\section{Simulation Results}
	\subsection{Simulation Parameters}
	For simulation, MIMO radar with simple case of two transmitters with four receivers is considered. To achieve waveform orthogonality, slow-time phase coding over FMCW waveform has been considered where outer hadamard matrix of order 2 from \eqref{hadamard} is utilized to implement the slow-time phase coding. The FMCW waveform parameters are listed in Table~\ref{tab:fmcw_radar}. The basic FMCW waveform parameters are configured to set the maximum range of 130 meters with a range resolution of 62.5 cm. The target scenario is assumed to consists of two targets where first target with RCS of -62 $dB_{sm}$, is set at 45 meters while the second target with RCS of 20 $dB_{sm}$, is set at 10 meters. The radar itself is static while mounted at a height of 2 meters from the ground. The frequency of operation is set in the W-band as various hardware devices supporting ST-PC FMCW based MIMO radars are commercially available in this band facilitating the verification of results through field experiments in the next section. The RCS of second target is kept low to demonstrate the robustness of proposed technique for different target scenarios.
	\subsection{Simulation Cases for Comparison}The comparison has been shown for three cases where the first case consists of targets at zero velocity while placed at boresight of radar whereas, the second case takes into consideration the effect of targets' doppler at boresight and finally, the third case demonstrates the robustness of proposed technique for moving targets scenario at random azimuth angles. For each case, three pulse compressed outputs were compared. The first plot (blue)  shows the traditional windowed (using Hanning window) matched filter output whereas, the second plot (green) shows the adaptive pulse compression output and finally, the third plot (red) shows the proposed technique with adaptive pulse compression on windowed fast-time samples after the application of doppler and angle processing. The input to these range compressed results is the decoded pulses as given in \eqref{decodingeqn}. No doppler compensation is implemented in these results to show the robustness of the proposed technique to doppler mismatch.    
	\subsubsection{Case 1: Two Static Targets at Radar Boresight}
	\begin{figure}[htbp]
		\centerline{\includegraphics[width=9cm]{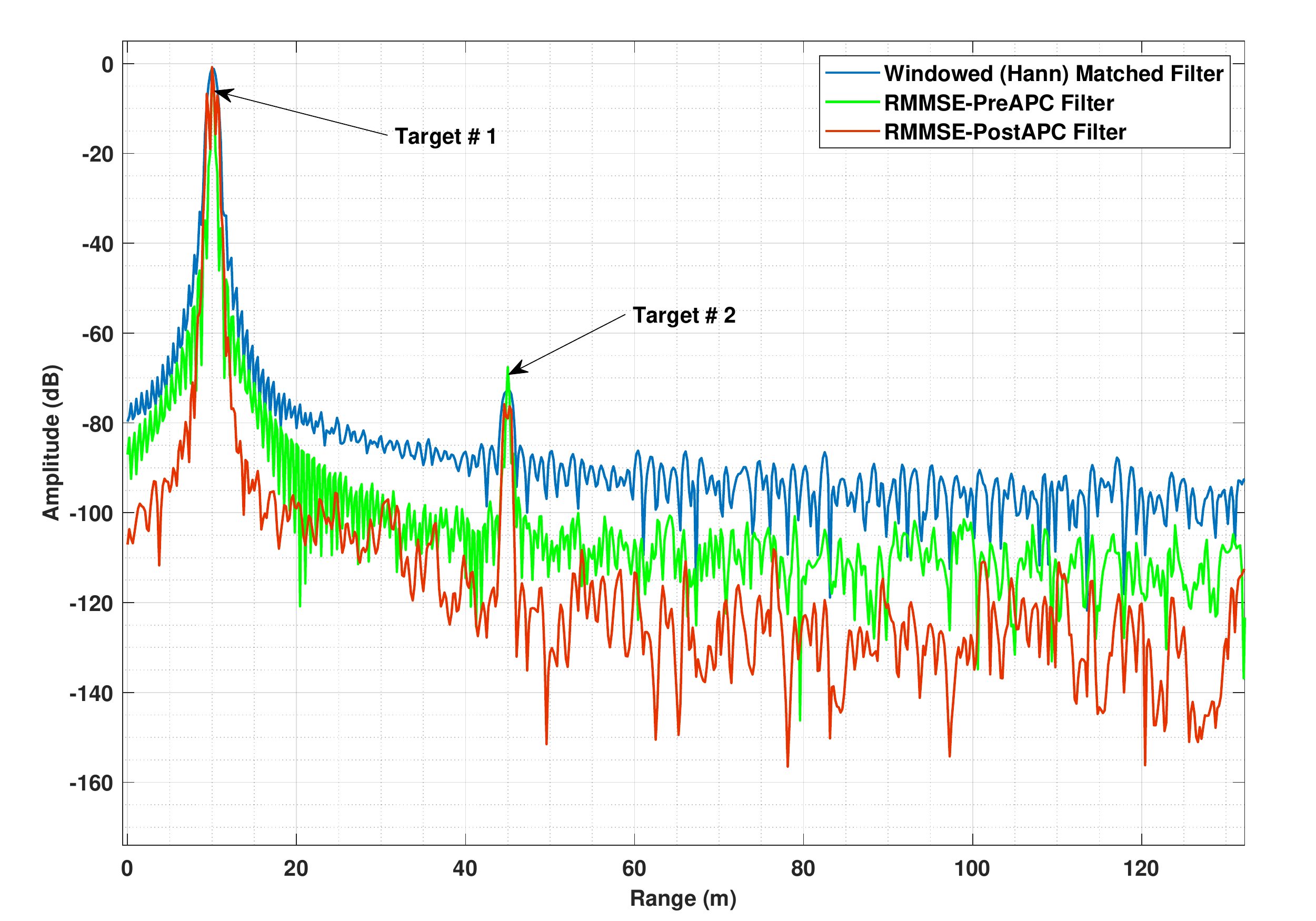}}
		\caption{Static targets at radar boresight with (a) Traditional pulse compression on Hanning windowed range samples (b) APC-RMMSE filter method (c) Proposed technique on doppler and angle processed data with Adaptive Pulse Compression (APC) on Hanning windowed range samples}
		\label{case1static}
	\end{figure}
	It can be observed from  Fig.~\ref{case1static} that all compared techniques provide significant range sidelobes reduction which subsequently reiterates that MIMO waveforms have been properly decoded with minimal residual phase distortions. Moreover, the baseline APC algorithm and the proposed APC algorithm provide approximately similar and improved Signal-to-Interference-plus-Noise-Ratio (SINR) as compared to the traditional Hanning windowed range response. 
	\subsubsection{Case 2: Two Moving Targets at Radar Boresight}
\begin{figure}[htbp]
		\centerline{\includegraphics[width=9cm]{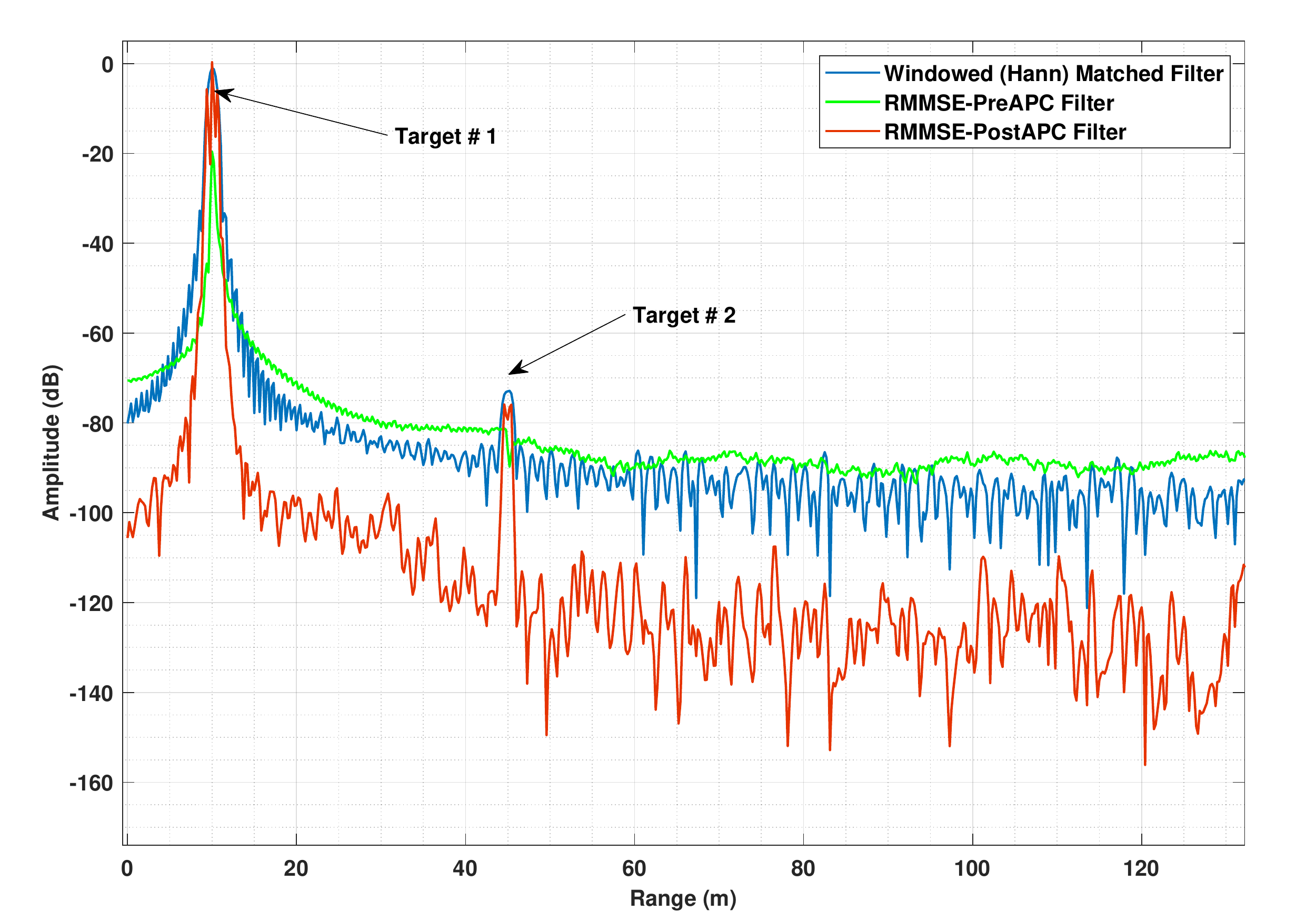}}
		\caption{Moving targets at radar boresight with (a) Traditional pulse compression on Hanning windowed range samples (b) APC-RMMSE filter method (c) Proposed technique on doppler and angle processed data with Adaptive Pulse Compression (APC) on Hanning windowed range samples}
		\label{case2movuncomp}
	\end{figure}

	The two targets are assumed to be moving at -20 m/s (Target 1) and 30 m/s (Target 2), respectively while spatially located at zero degree azimuth (boresight of radar antenna). From the results in  Fig.~\ref{case2movuncomp}, it can be observed that the range response of the baseline APC method has degraded largely due to the doppler phase distortion associated with doppler phase shift which subsequently does not provide proper decoding of received pulses. Additionally, the weaker target has gone undetected as the interference due to doppler phase distortion has masked the weaker target. Whereas, the range sidelobes response of traditional Hanning windowed as well as the proposed technique is unaffected. Thereby, it is noted that the application of the Hanning window mitigates the effect of phase distortion associated with doppler shifts of targets. Moreover, it can be observed that the sidelobes suppression using the proposed technique is much better as compared to the traditional windowing method while also providing improved SINR for the targets. In addition to this, the inherent advantage of narrower mainlobe associated with the baseline APC method is also visible in the proposed technique.
	\subsubsection{Case 3: Two Moving Targets at Radar Azimuth angles}
\begin{figure}[htbp]
		\centerline{\includegraphics[width=9cm]{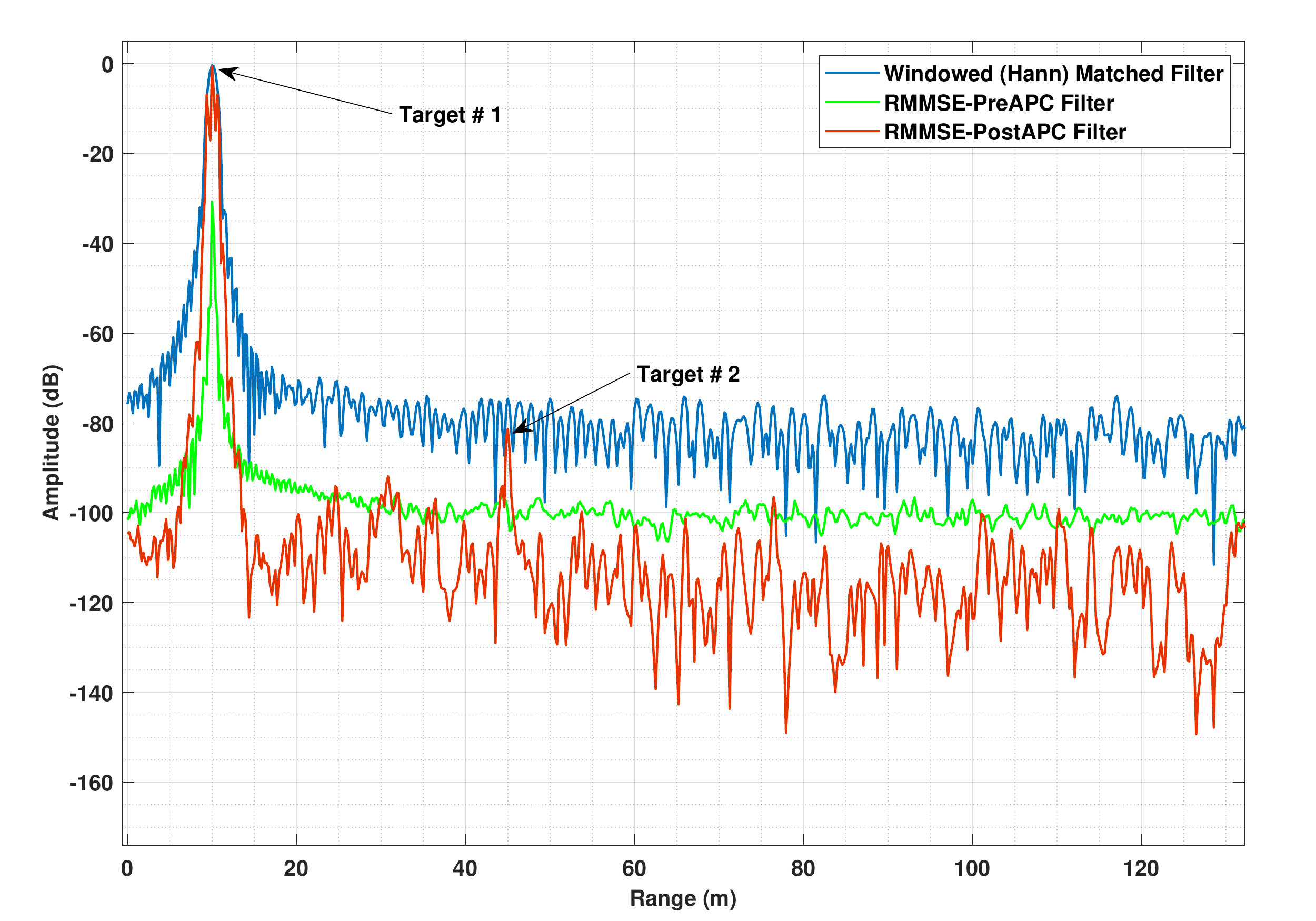}}
		\caption{Moving targets at random spatial angles with (a) Traditional pulse compression on Hanning windowed range samples (b) APC-RMMSE filter method (c) Proposed technique on doppler and angle processed data with Adaptive Pulse Compression (APC) on Hanning windowed range samples}
		\label{case3movang}
	\end{figure}

	For the final simulated case shown in  Fig.~\ref{case3movang}, the moving targets from the previous case are now spatially located at -5 and -10 degrees azimuth, respectively to assess the effect of phase distortion associated with the spatial location of targets in a ST-PC MIMO radar. This phase distortion is due to the extra path length covered by transmissions other than the first transmitter as mentioned in \eqref{mimoreceived}. The effect of combined phase distortions due to target velocities and spatial location can be observed in all the techniques. Again, the baseline APC method is unable to detect the weaker target while the SINR of the stronger target has also been reduced. In this case, even the traditional windowed technique is unable to detect the weaker target due to additional interference from phase distortions associated with the spatial location of targets thereby, causing an increase in its noise floor. Finally, the proposed technique can detect the weaker target though with a slight reduction in SINR whereas, the SINR of the strongest target stays approximately the same. Thereby, outperforming the other techniques in this scenario.    
	\section{Hardware Implementation}
\begin{figure}[htbp]
		\centering
		\includegraphics[width=6cm]{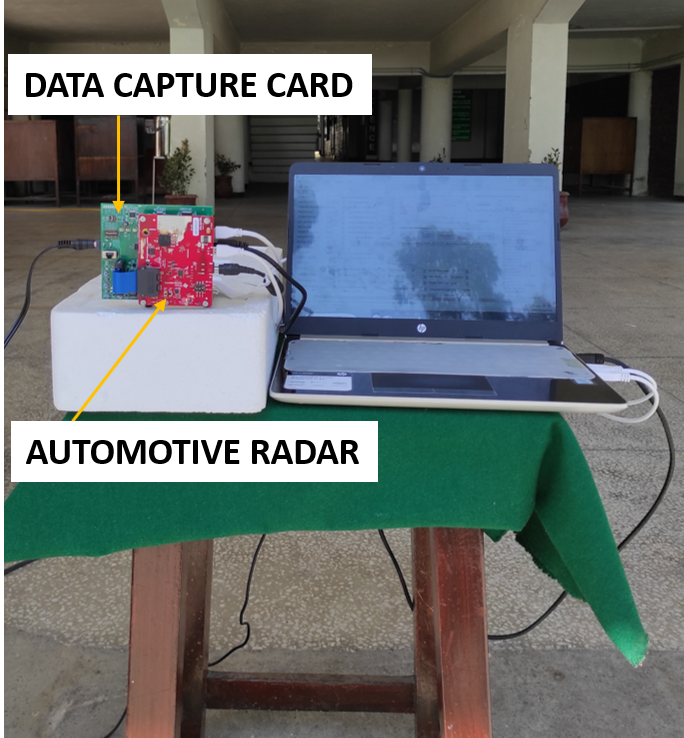}
		\caption{Automotive RF front-end installed with data capture device for post-processing}
		\label{hardware}
	\end{figure}

	The proposed methodology and RMMSE filter design technique is demonstrated using an FMCW MIMO radar in the automotive frequency band i.e. W Band (77 GHz). The MIMO radar system consists of two transmitters and four receivers with the basic FMCW waveform slope of 29.982MHz/$\mu$s. The waveform has a sweep duration of 60$\mu$s while operating from 77GHz to 78.792GHz. The radar and waveform specifications used for the hardware implementation of the proposed algorithm are listed in  Table~\ref{tab:hardware} with binary phase modulation (BPM) carried out on consecutive pulses to achieve MIMO waveform orthogonality through slow-time phase coding. A data capture device was used to capture the received pulses using the automotive MIMO RF front-end as shown in Fig.~\ref{hardware}, for subsequent post-processing on the raw data.	
	\begin{table}[h]
		\centering
		\caption{Radar and Waveform Design Parameters for Hardware Implementation}
		\begin{tabular}{|l|c|c|}
			\hline
			\textbf{Parameter} & \textbf{Variable} & \textbf{Value} \\ \hline
			Number of Transmitters & $N_T$ & 2 \\ \hline
			Number of Receivers & $N_R$ & 4 \\ \hline
			Start Frequency & $f_c$ & 77 GHz \\ \hline
			Waveform Bandwidth & $B$ & 1798.92 MHz \\ \hline
			Sweep Time & $T_p$ & 60$\mu$s \\ \hline
			Chirp Rate & $K$ & 29.982 MHz/$\mu$s \\ \hline
			ADC Sampling Frequency & $f_s$ & 10 MHz \\ \hline
			Number of Chirps in CPI & $N_p$ & 128 \\ \hline
			Near Range & $R_{near}$ & 0 \\ \hline
			Far Range & $R_{far}$ & 25.6 m \\ \hline
			Oversampling factor & $k$ & 3 \\ \hline
		\end{tabular}
		\label{tab:hardware}
	\end{table}
	\subsection{Target Scenario}
\begin{figure}[htbp]
		\centerline{\includegraphics[width=9cm]{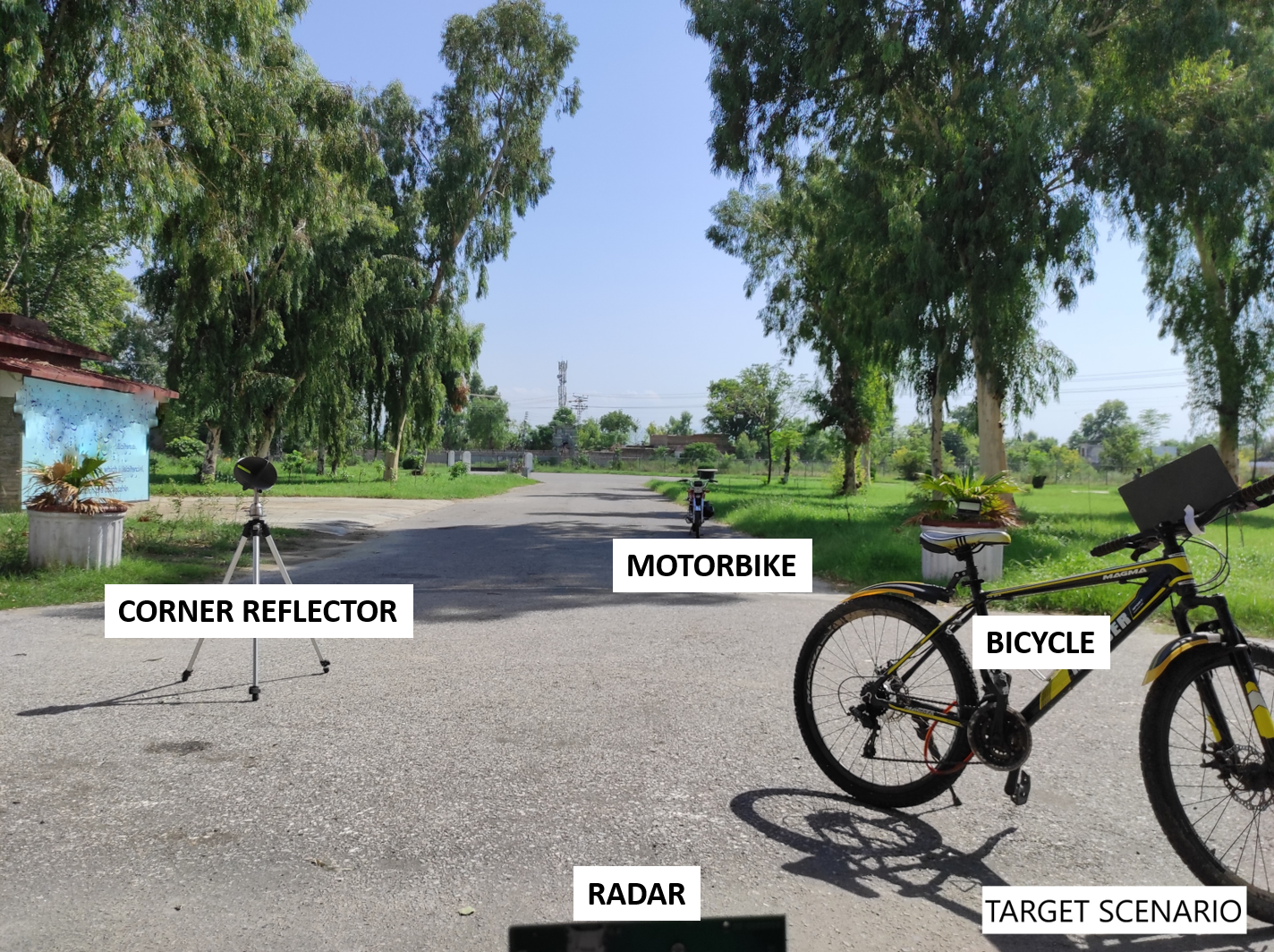}}
		\caption{Field experiment setup }
		\label{scenario}
	\end{figure}

\begin{figure}[htbp]
		\centerline{\includegraphics[width=9cm]{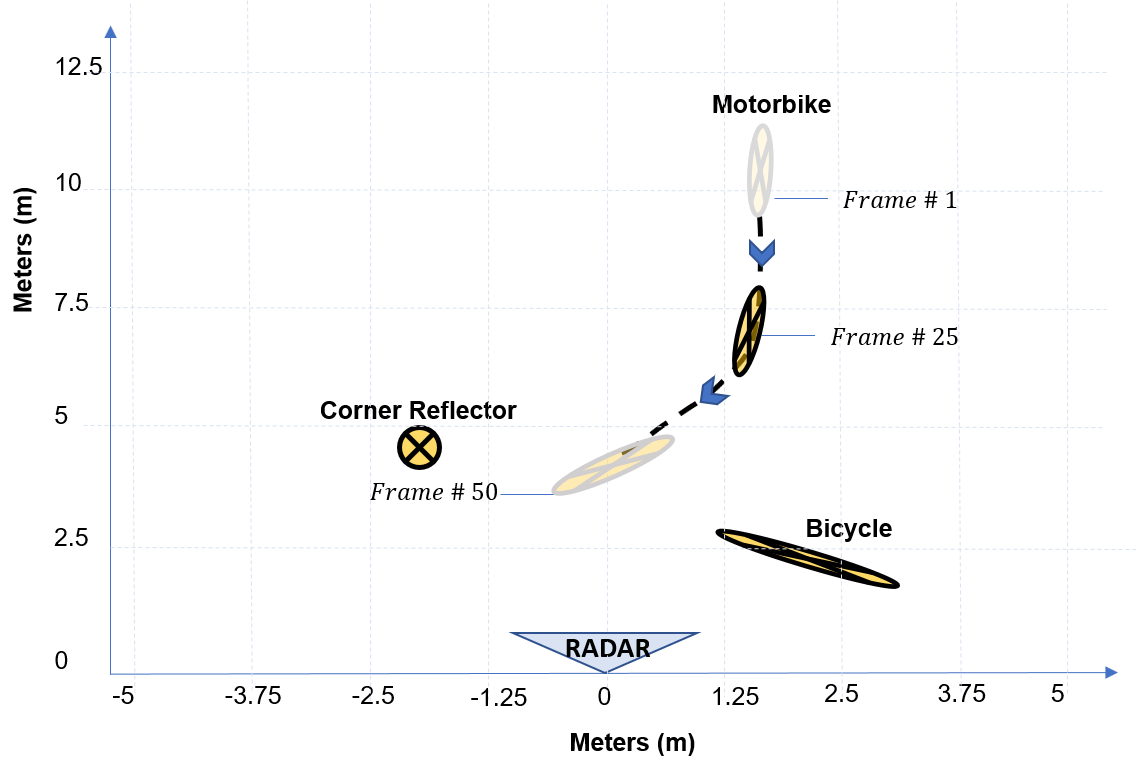}}
		\caption{Target scenario with two static targets and one moving target approaching the radar with increasing velocity}
		\label{scenarioblock}
	\end{figure}

\begin{figure}[htbp]
		\centerline{\includegraphics[width=9cm]{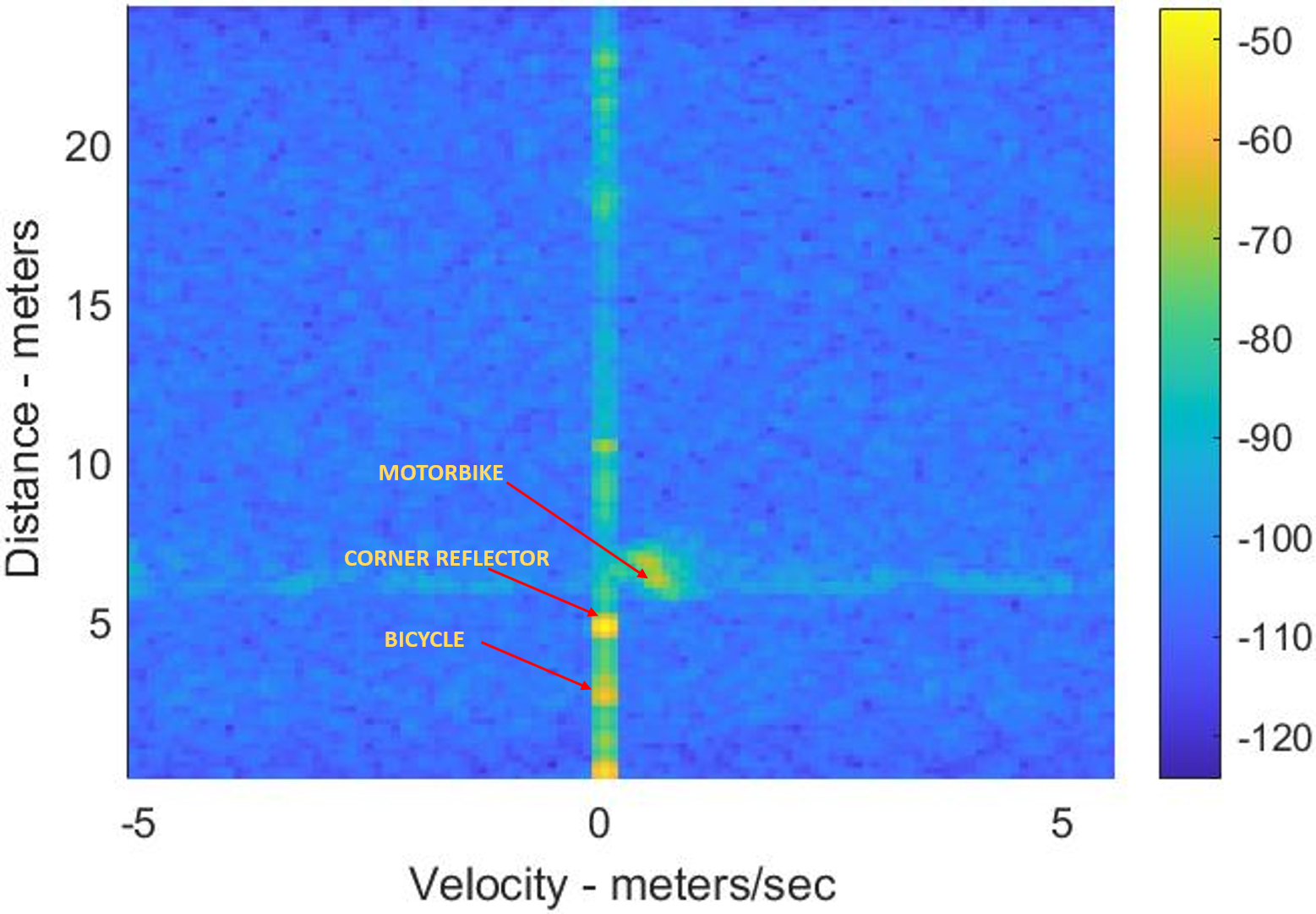}}
		\caption{Range-doppler response for frame 25 of captured returns showing two static targets (corner reflector and bicycle) and one moving target (motorbike)}
		\label{rangedoppler25}
	\end{figure}

	To show the robustness of the proposed algorithm, multiple targets were considered as shown in  Fig.~\ref{scenario}. As illustrated in Fig.~\ref{scenarioblock}, two targets i.e. corner reflector and bicycle were kept static at approximately 21 degrees and  42 degrees, respectively while the third target i.e. motorbike was made to follow a trajectory with increasing velocity while moving towards the radar. The bicycle and motorbike (at certain time instants) act as extended targets giving multiple range peaks as demonstrated later as well. As the bicycle is stationary, it provides a fixed extended range response with varying power levels starting from 1.1 to 3 meters. A total of 50 frames were captured to process the received returns from the target scenario whereas, the algorithm results were demonstrated for frame 25 captured returns (the corresponding range-doppler plot is shown in  Fig.~\ref{rangedoppler25}). A simple background subtraction technique was used to remove the static clutter other than the targets. Similar to the simulation scenarios, the proposed order of operations (\ref{oop}) has been implemented for our proposed algorithm (red) which were compared with the baseline APC algorithm (green) being implemented on received pulses using the traditional order of operations. For comparison with traditional sidelobes reduction technique, Hanning windowed range response (blue) on received pulses were also be compared. 
	\begin{figure}[htbp]
		\centerline{\includegraphics[width=9cm]{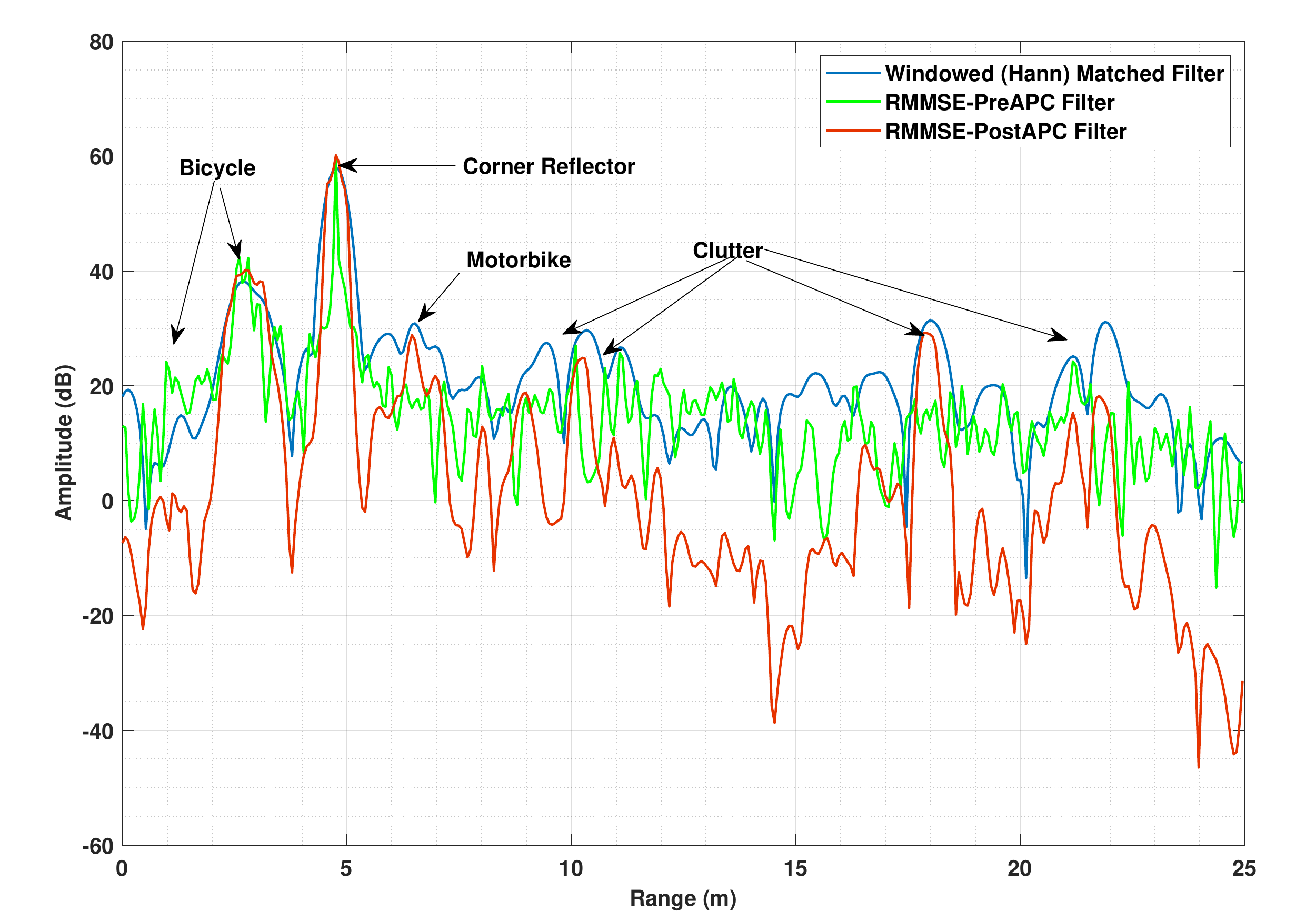}}
		\caption{Range response of frame 25 of captured returns with (a) Traditional pulse compression on Hanning windowed range samples (b) APC-RMMSE filter method (c) Proposed technique on doppler and angle processed data with Adaptive Pulse Compression (APC) on Hanning windowed range samples}
		\label{hardwareresults}
	\end{figure}
	\subsection{Results and Discussions}
	With the captured data of frame 25, the results obtained after the application of the proposed methodology are shown in Fig.\ref{hardwareresults}. Although background subtraction technique has been implemented to remove the static clutter but, still some peaks due to trees, plants etc. are still observable across the captured range response shown in Fig.\ref{hardwareresults}. It can be observed that the baseline APC algorithm (green) provides good target detection with considerable sidelobes reduction while providing narrower mainlobe width when applied to static targets i.e. corner reflector and bicycle. However, it is unable to detect the complete range response associated with bicycle and only provides peaks between 2.2 and 3 meters. Moreover, the baseline APC algorithm is also unable to detect the moving target (motorbike moving at 0.5 m/s) while positioned at approximately 15 degrees azimuth angle.  The results obtained for the baseline APC algorithm are in line with the simulated response where it was likewise shown in Fig.\ref{case3movang} that APC algorithm when applied directly to improperly decoded returns i.e. without doppler compensation, gives the degraded response.\\
	The Hanning windowed (blue) like the baseline APC algorithm also provides detections of static targets while providing much better SINR related to these peaks. It is also unable to detect the extended range response for the bicycle due to increased spatial phase distortion associated with the front wheel of the bicycle. However, it is also unable to detect the moving target (motorbike) which also contains an additional spatial phase distortion as it is placed at approximately 15 degrees for the captured time instant. Keeping in view the results of the captured frame, it is observed that these results verify the observations carried out for the simulated response shown in Fig.~\ref{case3movang} where weaker targets in improperly decoded returns with spatial phase shifts were undetected using the Hanning windowed range response.\\ 
	On the other hand, the proposed algorithm (red) for the captured returns of frame 25 can detect all three targets (static and moving) without the application of any phase compensation associated with the arbitrary phase distortions which subsequently results in improper decoding of received returns. Moreover, it provides greater reduction in sidelobes level ranging from 10 to 32 dB thereby, providing much better SINR for the detected targets. As a result, the moving target, which is otherwise undetected using the baseline APC algorithm and traditional windowing method, has been successfully detected while providing extended peak range response between 5.5 to 7 meters. In addition to this, the two peaks for the extended target (bicycle) are clearly observable giving an accurate length of the target starting from 1.2 to 3 meters, therefore, any detection algorithm such as Cell-Averaging Constant False Alarm Rate (CA-CFAR) will be able to detect the peaks associated with all the targets.\\
	The proposed algorithm as demonstrated through results (Fig.~\ref{hardwareresults}) from captured returns, provide robust performance in the presence of phase distortions due to doppler and spatial location of targets. Although doppler compensation may be applied at a later stage for accurate angle estimation by applying for doppler compensation on the detected range and doppler bins of the detected targets.  Henceforth, the proposed algorithm can detect the masked targets even with improper decoding of received returns in the presence of doppler and spatial phase shifts by iterative minimization of interference associated with autocorrelation and cross-correlation range sidelobes. 
	
	\subsection{Quantitative Analysis}
	In this section, detailed performance comparison of proposed filter is presented against other discussed filters in terms of weighted amplitude differential and moving average statistics. First, we consider the amplitude (\(X\)) of captured returns subject to particular filter as a random variable. The raw amplitude differentials between proposed filtered received signal and other filtered responses (\(\delta\lvert_{x_p,x_f}=X_p-X_f\)) are also random variables which have positive value at target locations, owing to processing gain of Post-APC technique. Whereas, negative values of these differentials can be associated to noise suppression attribute of proposed filter in comparison to other filters. However, raw amplitude differential is prone to be affected by the nominal value of received signal as we can observe large amplitude variation in any of the filtered signal in Fig. \ref{hardwareresults}. 
	
	Therefore, a weighted amplitude differential (\(\Delta\lvert_{x_p,x_f}\)) is introduced which is subjected to a factor of distance of amplitude of proposed filtered signal from its mean (\(\mu_{p}\)) in terms of standard deviation (\(\sigma_{p}\)) as given below.
	
	\begin{equation*}
		\centering
		\Delta \lvert_{x_p,x_f}=\left\lvert\dfrac{\sqrt{x^2_{p}-\mu^2_{p}}}{\sigma_{p}}\right\rvert* \delta\lvert_{x_p,x_f} 
	\end{equation*}
	
\begin{figure}[ht]
		\centering
		\adjincludegraphics[width=\columnwidth,trim={{0.07\width} {0.22\height} {0.07\width} {0.22\height}},clip]{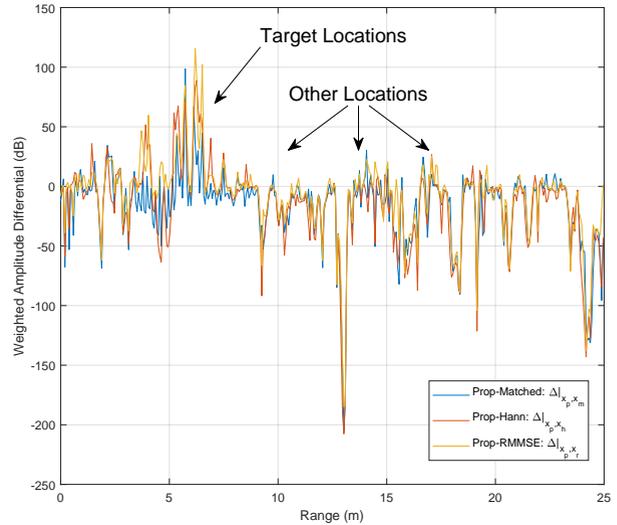}
		\caption{Weighted amplitude differential of proposed filter with other filters}
		\label{fig:w_diff}
	\end{figure}

	The effectiveness of proposed filter can be better visualized in Fig. \ref{fig:w_diff} as weight amplitude differential provide realistic performance analysis in comparison to matched, Hann and conventional RMMSE filter responses (\(\Delta\lvert_{x_p,x_m},\Delta\lvert_{x_p,x_h}\,\&\,\Delta\lvert_{x_p,x_r}\)), especially at target locations. Table \ref{tab:weighted_amp} shows the quantitative analysis of mean values of weighted amplitude differential at target and non-target locations. It can be clearly noticed that at target locations, the proposed filter provides considerable processing gain at these instances, maximizing it in comparison to RMMSE filter. At other non-target locations, the proposed filter provide significant noise suppression. 
	
	\begin{table}[h]
		\centering
		\begin{tabularx}{0.95\columnwidth} { 
				| >{\centering\arraybackslash}X 
				| >{\centering\arraybackslash}X
				| >{\centering\arraybackslash}X
				| >{\centering\arraybackslash}X|}
			\hline
			\multicolumn{4}{|c|}{\textbf{Weighted Amplitude Differential}}\\
			\hline
			\hline
			\textbf{Parameter} &  \textbf{Proposed-Matched}  & \textbf{Proposed-Hann}  & \textbf{Proposed-RMMS} \\
			
			\hline
			\(\mu_{\Delta^{'}}\) (dBm & 11.93 & 29.48 & 33.89\\
			Targets)& & & \\
			\hline
			\(\mu_{\Delta^{'}}\) (dBm & -20.77 & -22.69 & -15.80\\
			Other locations)& & & \\
			\hline
			
		\end{tabularx}\\
		\caption{Quantitative Analysis of Weighted Amplitude Differential}
		\label{tab:weighted_amp}
	\end{table}
	\par
	Moving statistics can provide deep insight of filter responses on small variations of range. These statistics can be considered as measure of how effectively targets are distinguished from clutter and noise is suppressed where targets are not present. For these calculations, range bins of size 0.325 meters (\(\Delta r\)) have been regarded as window size for moving statistics of capture returns. Mean values of these moving metrics have been compared. In case of discrete sampled measurements, \(k\) (5 samples) returns are observed within distance \(\Delta r\) and have moving average \(\Bar{x}_n\) represented as:
	
	\begin{subequations}
		\centering
		\begin{align}
			\Bar{x}_n&=\dfrac{1}{k}\sum^{n}_{i=n-k+1}x_i\\
			\mu_{\Bar{x}}&=\dfrac{1}{k N}\sum^N_{n=1}\sum^{n}_{i=n-k+1}x_i \label{eq:mean_mov_avg}
		\end{align}
	\end{subequations}
	The mean of this moving average (\(\mu_{\Bar{x}}\)) in (\ref{eq:mean_mov_avg}) shows the nominal amplitude levels while focusing on effects of filter performance on small range bins. Larger values for target locations means that this filter has distinguishes targets more than other discussed filters. While, lower mean value at non-target locations for proposed filter relates to overall noise suppression over complete radial range in comparison to other filters. 
	
	\begin{table}[h]
		\centering
		\begin{tabularx}{0.95\columnwidth} { 
				| >{\centering\arraybackslash}X 
				| >{\centering\arraybackslash}X
				| >{\centering\arraybackslash}X
				| >{\centering\arraybackslash}X
				| >{\centering\arraybackslash}X|}
			\hline
			\multicolumn{5}{|c|}{\textbf{Moving Average Statistics}}\\
			\hline
			\hline
			\textbf{Parameter} &  \textbf{Proposed}  & \textbf{Matched}& \textbf{Hann}  & \textbf{RMMS} \\
			\hline
			\(\mu_{\Bar{x}}\) (dBm & -17.34& -18.42 & -23.82 & -27.31\\
			Targets)& & & &\\
			\hline
			\(\mu_{\Bar{x}}\) (dBm & -45.17 & -37.22 & -36.22 & -39.75\\
			Other locations)& & & &\\
			\hline
		\end{tabularx}
		\caption{Quantitative Analysis of Moving Average over 0.325m range bins}
		\label{tab:my_label3}
	\end{table}
	
	The measure of variation in successive captured returns over small range bin is given by moving standard deviation \(\Bar{\sigma}_n\):
	
	\begin{subequations}
		\centering
		\begin{align}
			\Bar{\sigma}_n&=\sqrt{\dfrac{1}{k}\sum^{n}_{i=n-k+1}(x_i-\Bar{x}_n)^2}\\
			\mu_{\Bar{\sigma}}&=\dfrac{1}{\sqrt{k}N}\sum^N_{n=1} \sqrt{\sum^{n}_{i=n-k+1}(x_i-\Bar{x}_n)^2} \label{eq:mean_mov_std}
		\end{align}
	\end{subequations}
	
	The mean of this standard deviation (\(\mu_{\Bar{\sigma}}\)) in (\ref{eq:mean_mov_std}) encompasses filter response to targets considering the small variations over individual range bins. Larger deviation represents that proposed filter is more responsive to target captured returns and immediately suppresses noise levels around target peaks. Amplitude deviation is irrelevant at locations where targets are not present as the range response varies randomly.

	\begin{table}[h]
		\centering
		\begin{tabularx}{0.95\columnwidth} { 
				| >{\centering\arraybackslash}X 
				| >{\centering\arraybackslash}X
				| >{\centering\arraybackslash}X
				| >{\centering\arraybackslash}X
				| >{\centering\arraybackslash}X|}
			\hline
			\multicolumn{5}{|c|}{\textbf{Moving Standard Deviation Statistics}}\\
			\hline
			\hline
			\textbf{Parameter} &  \textbf{Proposed}  & \textbf{Matched}& \textbf{Hann}  & \textbf{RMMS} \\
			\hline
			\(\mu_{\Bar{\sigma}}\) (dBm) & 5.36 & 5.15 & 3.25 & 4.57\\
			\hline
			\(\mu_{\Bar{\sigma}}\) & 10.09 & 8.22 & 7.36 & 8.10\\
			(Targets)& & & &\\
			\hline
			\(\mu_{\Bar{\sigma}}\) & 4.87 & 5.08 & 2.72 & 4.30\\
			(Other locations)& & & &\\
			\hline
		\end{tabularx}
		\caption{Quantitative Analysis of Moving Standard Deviation over 0.325m range bins}
		\label{tab:my_label4}
	\end{table}

	\section{Conclusion}
	Mismatched filters based on Adaptive Pulse Compression (APC) can be used for suppression of autocorrelation as well as cross-correlation sidelobes associated with MIMO waveforms. We proposed a methodology with re-adjusted order of operations due to the use of slow-time phase coding to achieve orthogonal waveforms for FMCW based MIMO radar. For proper decoding, doppler processing was performed first to identify the doppler bins of the detected targets for phase compensation. Angle processing is then implemented which, combined with doppler processing, enhances the dynamic range of the received pulses thereby, providing greater sidelobes reduction. We then formulated the baseline RMMSE based APC algorithm originally designed for Single-Input (SI) radar systems to incorporate Multi waveform capability. The derived cost function related to MIMO radar waveforms was then used to find the closed-form expression similar to the one derived for SI Multi-static Adaptive Pulse Compression (MAPC) filter.\\
	The proposed approach was then tested using MATLAB simulations where it was shown that the proposed methodology provides better performance in terms of range sidelobes suppression in comparison to the baseline APC algorithm and traditional windowed range response. The performance was shown to be robust in the presence of various phase distortions due to doppler and spatial locations of targets while providing the detection capability of otherwise masked weaker target with target parameters closely matching a practical scenario. Field experiments were then performed to demonstrate the robustness of the proposed algorithm. Similar to simulations, it also provided superior performance in an open-air experiment while providing improved SINR for the detected targets while providing detections of otherwise weaker masked targets.   
	

	\nocite{*}
	\bibliographystyle{IEEEtran}
	\bibliography{Adaptive_MIMO_access}

	\begin{IEEEbiography}[{\includegraphics[width=1in,height=1.25in,clip,keepaspectratio]{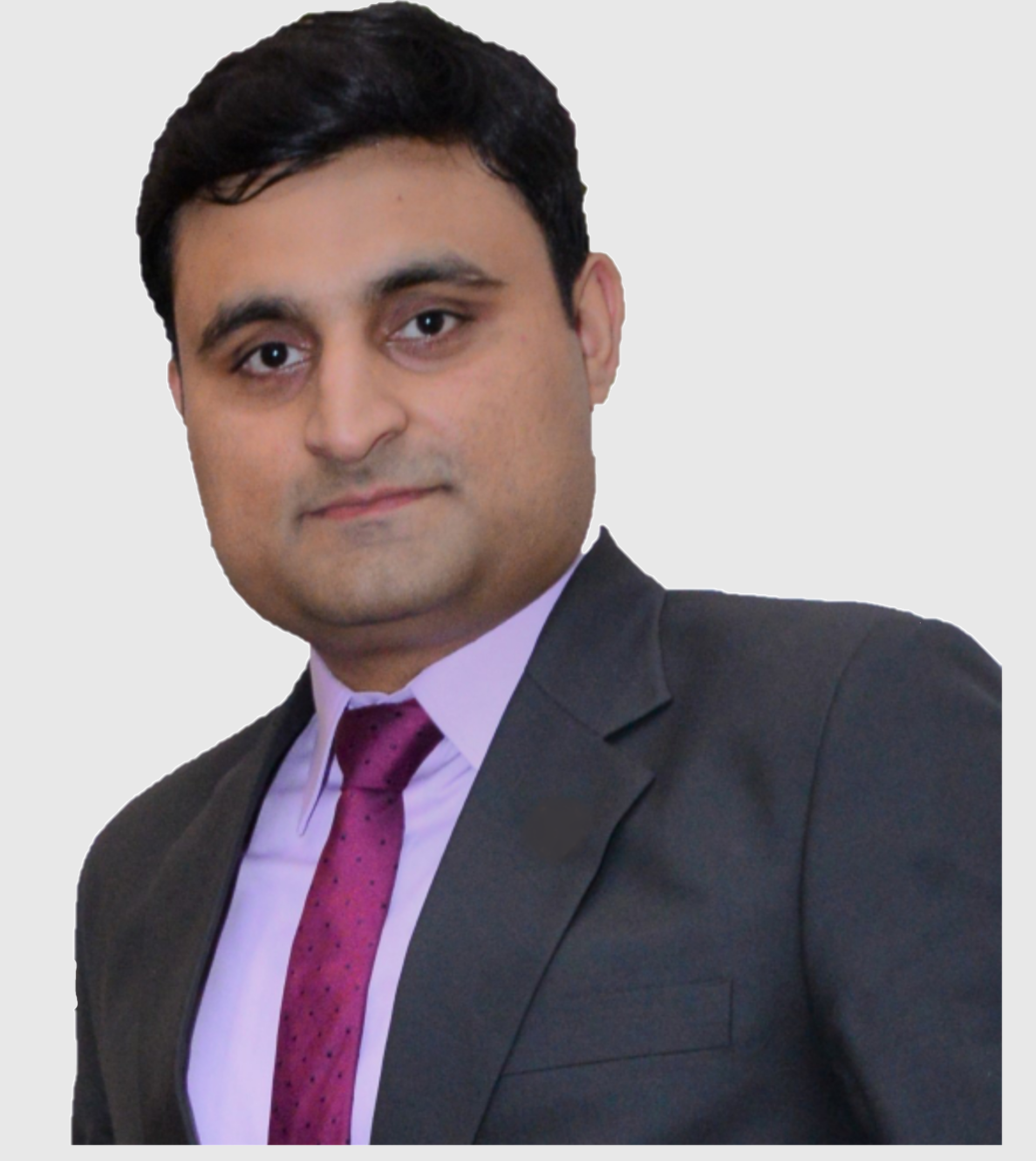}}]{Hamza Malik} 
		received Bachelors and Masters degrees in Avionics Engineering from National University of Sciences and Technology, Islamabad Pakistan, in 2013 and 2020, respectively.\\
		
		His research interests include radar signal processing, automotive radars, MIMO waveforms and embedded hardware.
	\end{IEEEbiography}
	
	\begin{IEEEbiography}[{\includegraphics[width=1in,height=1.25in,clip,keepaspectratio]{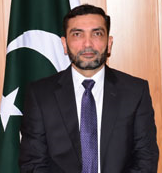}}]{Jehanzeb Burki} 
	received Masters and Ph.D. degrees in Electrical Engineering from Georgia Institute of Technology, Atlanta, GA, USA, in 2004 and 2008, respectively. He also holds Masters in Numerical and Applied Mathematics from School of Mathematics, Georgia Institute of Technology.\\
	
	He has been Professor and Head Avionics Engineering Department and Dean of College of Aeronautical Engineering (CAE), National University of Sciences and Technology (NUST), Pakistan. He as also been an adjunct faculty at Institute of Avionics and Aeronautics (IAA), Air University, Pakistan. He has over 25 years’ experience in basic and applied research in academia and industry. His research interests include radar signal processing, antenna design and engineering, RF and microwave circuit design, radar waveforms, remote sensing and wave propagation.
			
	\end{IEEEbiography}
	
	\begin{IEEEbiography}[{\includegraphics[width=1in,height=1.25in,clip,keepaspectratio]{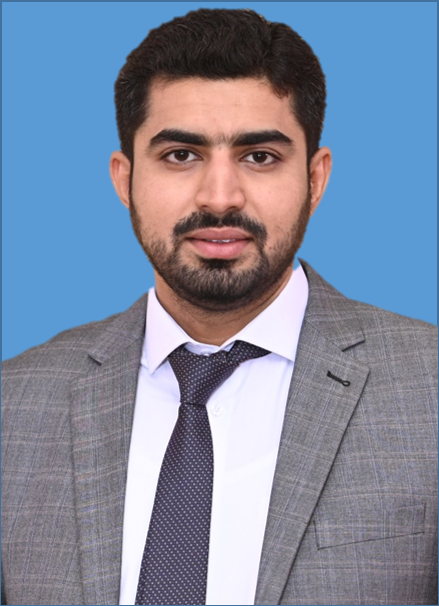}}]{Muhammad Zeeshan Mumtaz} 
	received Bachelors degree in Avionics Engineering from National University of Sciences and Technology, Islamabad Pakistan , in 2014 and Masters degree in Avionics Engineering in 2021 from Air University, Islamabad Pakistan.\\
	
	His research interests include signal processing and embedded hardware.
	\end{IEEEbiography}
	
	\EOD
	
\end{document}